\documentclass[conference,letterpaper]{IEEEtran}

\usepackage{times,amsfonts,amssymb,amsmath}
\usepackage{graphicx,url,bm,comment,subfigure}
\usepackage{epsfig}

\newcommand{\equaref}[1]{(\ref{eq:#1})}

\newcommand{\diff}{{\rm\,d}}

\newcommand{\Prob}
{
{\rm Pr}
}

\newtheorem{lemma}{\bf Lemma}

\newcommand{\ls}[1]
   {\dimen0=\fontdimen6\the\font
    \lineskip=#1\dimen0
    \advance\lineskip.5\fontdimen5\the\font
    \advance\lineskip-\dimen0
    \lineskiplimit=.9\lineskip
    \baselineskip=\lineskip
    \advance\baselineskip\dimen0
    \normallineskip\lineskip
    \normallineskiplimit\lineskiplimit
    \normalbaselineskip\baselineskip
    \ignorespaces
}

\newcommand{\tgifps}[3]{
\begin{figure}[htb]
\centering
\includegraphics[width=#1cm]{#2.ps}
\caption{#3\label{fig:#2}}
\vspace{-2mm}
\end{figure}
}

\newcommand{\ba}{\begin{array}}
\newcommand{\ea}{\end{array}}

\begin{document}


\begin{sloppypar}
\bibliographystyle{plain}
\nocite{*}

\title{Information-theoretic Capacity of Clustered Random Networks
}

\author{
\IEEEauthorblockN{Michele Garetto  \vspace{1mm}}
\IEEEauthorblockA{Universit\`{a} degli Studi di Torino\\
Corso Svizzera 185\\
10149 - Torino, Italy\\
Email: michele.garetto.at.unito.it}
\and
\IEEEauthorblockN{Alessandro Nordio  \vspace{1mm}}
\IEEEauthorblockA{IEIIT-CNR\\
Corso Duca degli Abruzzi 24\\
10129 - Torino, Italy \\
Email: alessandro.nordio.at.polito.it}
\and
\IEEEauthorblockN{Carla F. Chiasserini, Emilio Leonardi \vspace{1mm}}
\IEEEauthorblockA{Politecnico di Torino\\
Corso Duca degli Abruzzi 24\\
10129 - Torino, Italy \\
Email: lastname.at.polito.it}}

\maketitle
\begin{abstract}
\ls{0.90}
We analyze the capacity scaling laws of clustered ad hoc networks 
in which nodes are distributed according to a doubly stochastic shot-noise Cox
process. We identify five different operational regimes, and for each regime 
we devise a communication strategy that allows to achieve  a throughput  to within
a poly-logarithmic factor (in the number of nodes) of the maximum 
theoretical capacity.
\end{abstract}

\ls{0.90}

%

\vspace{-0cm}
\section{Introduction and related work}
The capacity of ad hoc wireless networks has been traditionally 
studied considering single-user communication schemes 
over point-to-point links \cite{Gupta-Kumar}.
Only recently \cite{bibbiatse,Leveque-Tse-four-reg,shah}, 
information-theoretic scaling laws of ad hoc networks have been 
investigated, showing that multi-user cooperative schemes 
can achieve much better performance than traditional
single-user schemes, especially in the low power attenuation regime. 

In this paper, we follow the stream of work \cite{bibbiatse,Leveque-Tse-four-reg,shah},
analyzing the information-theoretic capacity of clustered random networks
containing significant inhomogeneities in the node spatial distribution.
In particular, we consider nodes distributed according to 
a doubly stochastic Shot-Noise Cox Process (SNCP) over a square region
whose edge size can scale with the number of nodes. 

We provide both information-theoretic upper-bounds to the achievable capacity 
and constructive lower-bounds, which are asymptotically tight to within a poly-log factor
(in the number of nodes). Our study reveals the emergence of five 
operational regimes, in which different communication schemes combined with proper    
scheduling/routing strategies must be adopted to achieve the system capacity.

With respect to previous work, we provide several contributions.
First, the analysis in \cite{bibbiatse,Leveque-Tse-four-reg} is limited to networks 
in which nodes are uniformly distributed. In contrast, our complete
characterization of the network capacity achievable under the SNCP model
extends the analysis to a much broader class of network topologies (including
the uniform distribution as a special case), which can take into account 
the clustering behavior usually found in real systems. 

Second, the impact of inhomogeneities in the node spatial distribution has been
first investigated in \cite{shah}, where authors have found that
for small path-loss exponents (i.e., $\alpha \in (2,3]$) the capacity
does not depend on how nodes are placed over the area. Instead,
they show that capacity is significantly affected by the network topology
for large path-loss exponents (i.e., $\alpha > 3$). However, their
characterization of the capacity achievable for 
large path-loss exponents is limited to the case of adversarial
node placement under a deterministic (given) degree of network regularity. Moreover,
they impose a minimum separation constraint between the nodes which 
does not allow to introduce highly dense clusters over the area.
At last, the analysis in \cite{shah} is limited to the case 
of extended networks (i.e., networks whose area grows linearly 
with the number of nodes\footnote{In \cite{Leveque-Tse-four-reg}
authors recognized the importance of letting the network
area scale with the number of nodes in a general way, as this gives
rise to a richer set of operational regimes.}).

Third, our constructive lower bounds require to employ novel 
scheduling/routing strategies in combination to existing
cooperative communication schemes. Such strategies 
represent an important contribution in themselves, as they
could be adopted to cope with the nodes spatial 
inhomogeneity in more general topologies which cannot be 
described by the SNCP model considered here.
 
At last we emphasize that this work extends \cite{upper,lower},
where we have analyzed the capacity of networks in which nodes 
are distributed according to a SNCP model, but considering single-user
communication schemes only (i.e., traditional point-to-point links).

\section{System Assumptions and Notation}\label{sec:notation}

\subsection{\bf Network Topology}
We consider a network composed of a random number $N$ of nodes
(being  $E[N]= n$) distributed over a square region $\cal O$ of edge length
$L$, where $L$ takes units of distance. 
The network physical extension $L$ scales with the average number
of nodes, since this is expected to occur in many growing systems.
Throughout this work we will assume that $L=n^\gamma$, with $ \gamma \ge 0$.
To avoid border effects, we consider wrap-around conditions at the network
edges (i.e., the network area is assumed to be the surface of a bi-dimensional Torus).


The clustering behavior of large scale systems is taken into account
assuming that nodes are placed according to a shot-noise Cox
process (SNCP). An SNCP \cite{cox} over an area $\mathcal{O}$  can be conveniently described by the following
construction.  We first specify a homogeneous Poisson point process $\cal C$ of cluster centres,
whose positions are denoted by ${\bf C} = \{c_j\}_{j=1}^{M}$,
where $M$ is a random number with average $E[M]=m$.
Each centre point $c_j$ in turn independently generates   a point process of nodes whose intensity
at $\xi$ is given by $q k(c_j,\xi)$, where $q \in (0,\infty)$ and  
$k(c_j,\cdot)=k(\|\xi - c_j\|)$
is a rotationally invariant 
dispersion density function, also called kernel, or shot; i.e., $k(c_j,\cdot)$
depends only on the euclidean distance $\|\xi - c_j\|$ of point $\xi$
from the cluster centre $c_j$. 

Moreover we assume that
 $k(\|\xi - c_j\|)$ is a non-negative, non-increasing,  bounded and continuous function,
 whose integral $\int_{\cal O} k(c_j,\xi) \diff \xi$
over the entire network area is finite and equal to 1.
In practice, the kernels considered in our work can be specified by
first defining a non-negative, non-increasing continuous function $s(\rho)$ such that  $\int_0^\infty \rho~s(\rho) \diff \rho < \infty$
and then normalizing it over the network area $\cal O$:
$$ k(c_j,\xi) = \frac{s(\|\xi-c_j\|)}{\int_{\cal O} s(\|\zeta-c_j\|) \diff \zeta}$$
Notice that, in order to have finite integral over increasing network areas, 
functions $s(\rho)$ must be $o(\rho^{-2})$, i.e., they must have a tail
that decays with the distance faster than quadratically.
In the following, we will be especially interested in functions $s(\rho)$ whose tail
decays as a power-law: \begin{equation}\label{delta}s(\rho)=\min(1,\rho^{-\delta})\ \textrm{ for } \delta > 2,\end{equation} although our results apply to more general shapes as well.

At last, in our asymptotic analysis we can neglect the normalizing 
factor $\int_{\cal O} s(\|\zeta-c_j\|) \diff \zeta = \Theta(1)$. 

Under the above assumptions on the kernel shape, quantity $q$ equals
the average number of nodes generated by  each cluster centre (all cluster centres generate on average the same number
of nodes).
In our work, we let $q$ scale with $n$ as well (clusters are expected to grow
in size as the number of nodes increases).
This is achieved assuming that the average number of cluster centres
scales as $m = n^\nu$, with $\nu \in (0,1)$. Consequently,
the number of nodes per cluster scales as $q =  n^{1-\nu}$.

The overall node process $\cal N$ is then given by the superposition of the individual processes
generated by the cluster centres. The local intensity at $\xi \in \mathcal{O}$ of the resulting SNCP is
$$ \Phi(\xi) = \sum_{j =1}^{M}  q \,k(\|\xi - c_j\|).$$

Notice that $\Phi(\xi)$ is a random field, in the sense that,
conditionally over all $(c_j)$, the node process $\cal N$ is an
(inhomogeneous) Poisson point process with intensity function
$\Phi$. We denote by ${\bf X}=\{X_i\}_{i=1}^{N}$ the collection of
nodes positions in a given realization of the SNCP. 


Let $d_c = L/\sqrt{m} = n^{\gamma-\nu/2}$ be the typical distance
between cluster centres. More precisely, $d_c$ is the edge of the square
where the expected number of cluster centres falling in it equals 1.
We call \begin{list}{}{}
\item{\em cluster-dense} condition the case $\gamma < \nu/2$, in which
$d_c$ tends to zero an $n$ increases;
\item{ {\em cluster-sparse}} condition the case
$\gamma > \nu/2$, in which $d_c$ tends to infinity an $n$ increases.
\end{list}

\begin{figure}
     \centering
     \subfigure[SNCP with $\nu = 0.6$.]{
          \label{subfig:example1}
          \includegraphics[height=1.65in,angle=0]{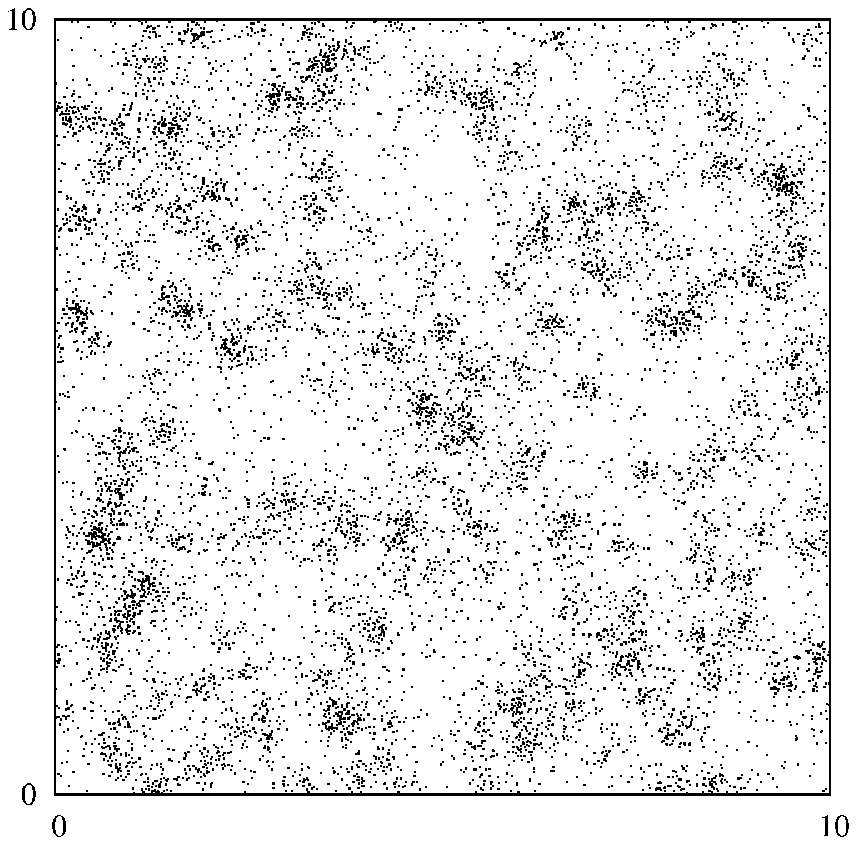}}
     \hfill{}
     \subfigure[SNCP with $\nu = 0.3$.]{
          \label{subfig:example3}
          \includegraphics[height=1.65in,angle=0]{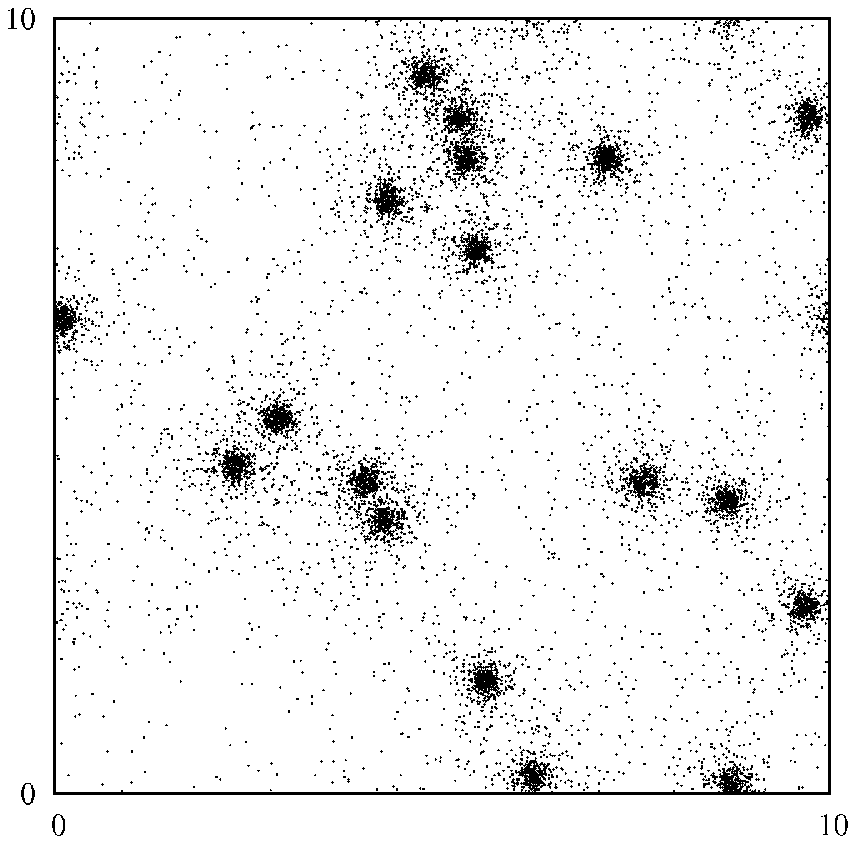}}
     \caption{Examples of topologies with $n = 10,000$ nodes distributed over the 
     square $10\,\times\,10$ ($\gamma=0.25$). In both cases $s(\rho) \sim \rho^{-2.5}$.}
    \label{fig:geom}
    \vspace{-2mm}
\end{figure}

Figure \ref{fig:geom} shows two examples 
of topologies generated by our SNCP, in the case of $n=10,000$ and $\gamma = 0.25$.
In both cases we have assumed  $s(\rho)=\min(1,\rho^{-2.5})$. The 
topology in Figure \ref{subfig:example1}
has been obtained with $\nu = 0.6$, hence it satisfies the {\em cluster-dense} 
condition ($\gamma < \nu/2$).
The topology in Figure \ref{subfig:example3} corresponds to $\nu = 0.3$, 
and provides an example of the {\em cluster-sparse} condition ($\gamma > \nu/2$).

Recall that the local intensity of nodes at point $\xi$  can be written as $\Phi(\xi)=\sum_j q\, k(c_j,\xi)$.
We define the quantities:  
$\overline{\Phi}=\sup_{\xi \in \cal O} \Phi(\xi)$ and $ \underline{\Phi}=\inf_{\xi \in\cal  O} \Phi(\xi)$.
The following lemma, proven in \cite{upper}, characterizes the asymptotic behavior of $\overline{\Phi}$ and $\underline{\Phi}$:
\begin{lemma}\label{teo:dens}
Consider nodes distributed according to the SNCP. Let $\eta(m)= d_c \sqrt{\log{m}}$.
If \mbox{$\eta(m) = o(1)$}, it is possible  to find two positive constants
$g_1,G_1$ with $g_1 < G_1$ such that $\forall \xi \in {\cal O}$
\begin{equation}\label{eq:denserg}
 g_1 \dfrac{n}{L^2}< \Phi(\xi)<G_1\dfrac{n}{L^2} \qquad  \text{w.h.p.\footnotemark}
\end{equation}
\footnotetext{Throughout the paper, 
we adopt the terminology  `with high probability' (w.h.p.)  to indicate 
events/properties  that  occur with a probability \mbox{$p= 1 - O(\frac{1}{n})$}, when $n \to \infty$.}
When \mbox{$\eta(m) = \Omega(1)$}, it is possible to find two positive constants
$g_2,G_2$, such that, w.h.p., \mbox{$\underline{\Phi}>g_2 q\,{\log m}\,s(d_c \sqrt{\log{m}})$}
and \mbox{$\overline{\Phi} <  G_2 q\,\log{m}$}.
\end{lemma}
The above result implies that $\underline{\Phi}=\Theta(\overline{\Phi})$ in the {\em cluster-dense} condition,  
i.e., when $\gamma < \nu/2$ (which implies  $d_c=o(1/\sqrt{\log m})$), whereas 
$\underline{\Phi}=o(\overline{\Phi})$ in the {\em cluster-sparse} condition,  i.e., when  
$\gamma > \nu/2$ (which implies $d_c=\omega(1)$). 

\vspace {-1mm}
\subsection{\bf Communication Model}
We use the same channel model as in \cite{bibbiatse, Leveque-Tse-four-reg, shah}.
Consider the generic time $t$, and let $V(t)$ be the set of nodes
transmitting at time $t$.
The signal received at time $t$ by a node $k$ is
$$ y_k[t] = \sum_{i \in V(t)\setminus \{k\}} h_{i,k}[t] x_i[t] + z_k[t] $$
where $x_i[t]$ is the signal emitted by node $i$, and $\{z_k[t]\}_{k,t}$ 
are white circularly symmetric Gaussian noise, independently and identically 
distributed (i.i.d.) with distribution ${\cal N}_{\mathbb C}(0,N_0)$ (with zero mean and variance $N_0$ per symbol). 
The complex baseband-equivalent channel gain $h_{i,k}[t]$ between $i$ and $k$
at time $t$ is
$$ h_{i,k}[t] = \sqrt{G} d_{ik}^{-\alpha/2} e^{j \theta_{ik}[t]} $$
where $G$ is a constant gain, $\alpha > 2$ is the path-loss exponent, and $\{\theta_{ik}[t]\}_{i,k}$
are i.i.d. random phases with uniform distribution in $[0,2\pi)$, which are assumed
to vary in a stationary ergodic manner over time (fast fading).
Moreover, $\{\theta_{ik}[t]\}_{i,k}$ and $\{d_{ik}\}_{i,k}$ are also assumed to be
independent, $\forall i,k$. We should mention that a recent work \cite{ahia} 
has put in discussion the validity of this assumption for very large $n$ in the low path-loss regime
$\alpha \in (2,3)$. However, the strong impact of the assumptions on the location of scatterers 
suggests that channel modelling in the low path-loss regime for very large networks is 
somewhat delicate and requires further investigation.

We assume that each node is source and destination of a single flow, 
and that the resulting $N$ flows (with $E[N]=n$)  are established at random without
any consideration of node locations. Let $\lambda(n)$ be the largest 
uniformly achievable rate of communications between sources
and destinations. The aggregate system capacity is $C(n) = n \lambda(n)$.
At last, we impose an average power constraint of $P$ on the transmissions
performed by each node, where $P$ is a constant. 

Table \ref{tab:system} summarizes the parameters of our model.
For the quantities that are allowed to scale with $n$ we have reported,
in the third column, the restrictions on the scaling exponent in $n$, i.e.,
the assumptions on $\log_n(\text{$<$parameter$>$})$.
\begin{table}
\centering
\caption{\label{tab:system} System parameters (\mbox{\rm n.a.} = not applicable)}
\vspace{-3mm}
\begin{tabular}{|l|l|c|}
\hline
\! Symbol \!  & \! Definition \! & \! scaling exponent \! \\
\hline
$L$   & edge length of network area & $\gamma \ge 0$ \\
$m$  &  average number of clusters & $0 < \nu < 1$ \\
$P$  & per-node power budget & $0$ \\
$\alpha$ & path-loss exponent & n.a. \\
$\delta$ & decay exponent of $s(\rho)$ & n.a. \\
$d_c$ & typical distance between cluster centres & $ \gamma - \nu/2$ \\
$q$ & average number of nodes per cluster & $1-\nu$ \\
\hline
\end{tabular}
\vspace{-2mm}
\end{table}
Note that $d_c$ and $q$ are not native parameters, since they are derived
from others, however we have included them in the table for
convenience.

\section{Summary of Results}\label{sec:results}
Similarly to previous work \cite{bibbiatse,Leveque-Tse-four-reg},  we express our results in terms of the 
scaling exponent $e_C$ of the network capacity, defined as,
$$ e_C =  \lim_{n \rightarrow \infty} \frac{\log C(n)}{\log n} $$
The scaling exponent allows to ignore all poly-logarithmic factors, i.e., factors which are 
$O(\log{n})^k$, for any finite $k$. Since our lower and upper bounds differ
at most by a poly-logarithmic factor, the corresponding scaling exponents
match, and we can claim that our characterization of the network capacity 
in terms of the scaling exponent is exact. 

Results are reported in Table \ref{tab:results}. The scaling exponent takes
different expressions as functions of the four system parameters $\{\alpha,\gamma,\delta,\nu\}$,
under the conditions specified in the third column of Table \ref{tab:results}.
In particular, we can distinguish five operational regimes, denoted by latin
numbers I,II,\ldots,V, as reported in the second column of the table\footnote{In the last two rows of the table,
the actual regime depends on which term prevails in the $\max(\cdot)$ expression
used in column one: we are in regime III if the first term is bigger than the second one.}.
\begin{table*}
\newcommand\T{\rule{0pt}{2.2ex}}
\newcommand\B{\rule[-1.3ex]{0pt}{0pt}}
\centering
\caption{Scaling exponent of network capacity \,\,\,$\beta = 1-\nu-\delta(\gamma - \nu/2)$.
\label{tab:results}}
\vspace{-3mm}
\begin{tabular}{|c|c|l|}
\hline
$e_C$ & regime & conditions \\
\hline
$1$   & I & $\alpha \gamma \le 1$ \T \\ 
$2 - \alpha \gamma$ & I & $\alpha \gamma > 1 \wedge \alpha \le 3$ \T \\
$\frac{\alpha -1 -\alpha\gamma}{\alpha-2}$ & II & $\alpha \gamma > 1 \wedge  \alpha > 3 \wedge \frac{1-2\gamma}{\alpha-2} \ge \gamma - \frac{\nu}{2}$ \T \\ 
$\max \big[ 2 - \alpha\gamma +  (\alpha-3) \frac{\nu}{2}, \gamma + \beta \frac{\alpha-1}{\alpha-2} \big]$ & III or IV & 
$\alpha \gamma > 1 \wedge  \alpha > 3 \wedge \frac{1-2\gamma}{\alpha-2} < \gamma - \frac{\nu}{2} \wedge \beta > 0$ \T \\
$\max \big[ 2 - \alpha\gamma + (\alpha-3) \frac{\nu}{2}, \gamma + \beta \frac{\alpha+1}{2} \big]$ & III or V & 
$\alpha \gamma > 1 \wedge  \alpha > 3 \wedge \frac{1-2\gamma}{\alpha-2} < \gamma - \frac{\nu}{2} \wedge \beta \le 0$ \T \\
\hline
\end{tabular}
\vspace{-2mm}
\end{table*}
It can be verified that $e_C$ varies with continuity in the four-dimensional space 
of parameters $\{\alpha,\gamma,\delta,\nu\}$. We observe that, under any regime, $e_C$ is a non-increasing
function of parameters $\{\alpha,\gamma,\delta\}$ and a non-decreasing function of $\nu$.
Figure \ref{fig:summary} provides a graphical representation of the results in Table \ref{tab:results}
for the particular case of $\nu = 0.3$ and $\delta = 2.5$.

\tgifps{8}{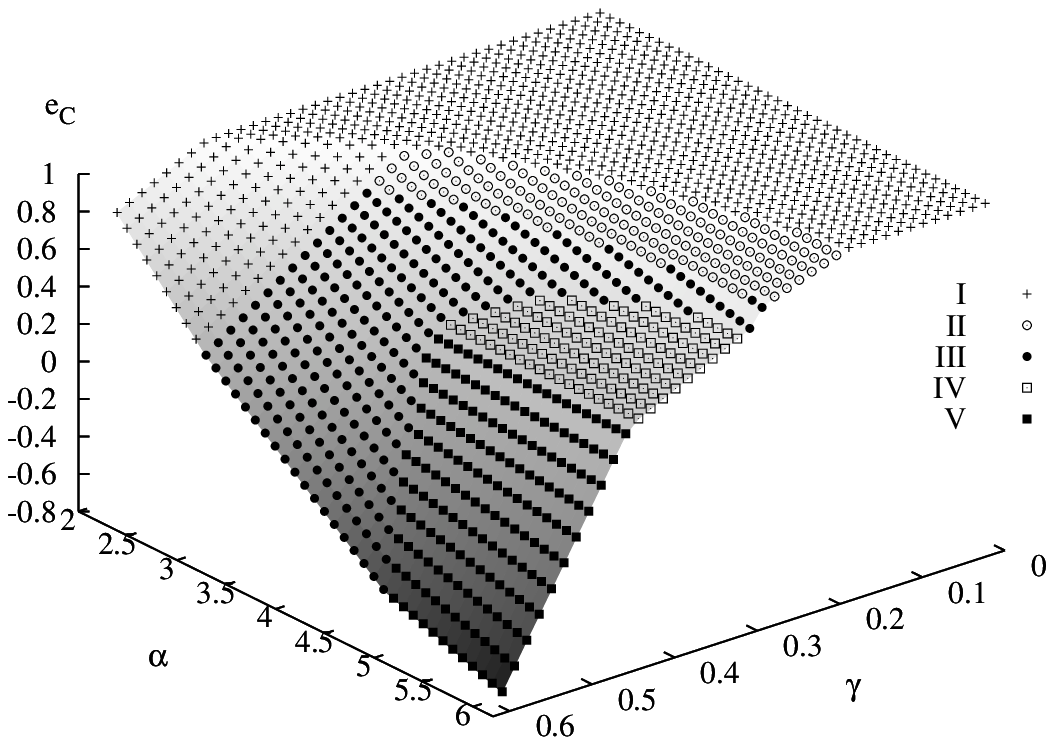}{Scaling exponent of network capacity as function of $\alpha$ and $\gamma$, for 
$\nu = 0.3$ and $\delta = 2.5$. Different marks are associated to the five possible regimes.}

\section{Upper Bounds}\label{sec:upper}
Upper-bounds are obtained extending the approach in \cite{bibbiatse,Leveque-Tse-four-reg,shah}, 
which is based on the computation of a bound to the information flow passing
through a cut that divides the network in two parts.

First, by leveraging percolative arguments (see \cite{upper}), it is possible to find a 
strip of width $\Delta$, with $\Delta=\Theta((q s(d_c)\log n)^{-1/2})$ in the {\em cluster-sparse}
condition (i.e., for $\gamma>\nu/2$) and $\Delta=\Theta(L/\sqrt{n})$
in the {\em cluster-dense} condition (i.e., for $\gamma<\nu/2$), which divides the network area in two parts, 
and satisfies the following properties: i) the considered strip is empty of nodes;
ii) every cluster centre lies at a distance greater than $g\,d_c$ from the strip,
for a sufficiently small constant $g$. 

Then, the information flow $\cal C({\mathcal S},{\mathcal D})$ from sources ${\mathcal S}$ on the left
to destinations ${\mathcal D}$ on the right can be bounded by the 
power transfer ${\cal P}_{{\mathcal S},{\mathcal D}}$ through the strip, 
as in \cite{bibbiatse,Leveque-Tse-four-reg,shah}. Let  $d_{ik}$ be the
euclidean distance between node $i$ and node $k$. According to \cite{shah}:\vspace{1mm}
\begin{equation}
 {\cal C}({\mathcal S},{\mathcal D}) \le b \,{\cal P}_{{\mathcal S},{\mathcal D} }= 
 b \sum_{i\in {\mathcal S},k\in {\mathcal D}} P\, d_{ik}^{-\alpha}
\label{upper-power}
\end{equation}
for any \vspace{-3mm}  
$$b > 4 \max \left( 1,\max_{k \in  {\mathcal D}} \sum_{i\in {\mathcal S}}
\frac{|h_{ik}|^2 }{ \sum_{h \in  {\mathcal D}}d_{ih }^{-\alpha} } \right)$$ 
 being, for every $k$,  $ \sum_{i\in {\mathcal S}}\left(
\frac{|h_{ik}|^2} { \sum_{h \in  {\mathcal D}}d_{ih }^{-\alpha} }\right) = O( \log^5 n)$.
 
To estimate ${\cal P}_{{\mathcal S},{\mathcal D} }$,   
the left and right domains are partitioned, respectively, into 
squarelets $\{A_k\}_k$ and $\{B_h\}_h$, obtaining: \vspace{1mm}
 \[
{\cal P}_{{\mathcal S},{\mathcal D} }= P \sum_{i\in {\mathcal S},k\in {\mathcal D}}
d_{ik}^{-\alpha} \le P \sum_h \sum_k \underline{d}_{hk}^{-\alpha}\overline{U}(A_k) \overline{U}(B_h)
 \]  
where $\underline{d}_{hk}$ is the minimum distance between points of  $A_k$
and points of $B_h$, while function $\overline{U}(A_k)$ ($\overline{U}(B_h)$)
 provides   an upper bound to the
number of nodes in $A_k$ ($B_h$). To obtain tight upper bounds the size of $A_k$ and
$B_h$ must be carefully chosen since, by increasing 
the size of $A_k$ and $B_h$, on the one hand we obtain tighter bounds for
$\overline{U}(A_k)$ and $\overline{U}(B_h)$; on the other, we
obtain looser bounds $\underline{d}_{hk}$ for the distance between
nodes in $A_k$ and nodes in $B_h$.

Furthermore, when $\Delta=o(1)$, a tighter bound can be obtained
applying the Hadamard inequality (see \cite{Leveque-Tse-four-reg}) 
to extract from the information flow the 
contribution of destinations receiving signals whose strength diverges. 
This contribution, associated to nodes in proximity of the cut, 
is in turn evaluated applying the Hadamard inequality iteratively, so as to split it
into the contributions associated to individual destinations
(which can be interpreted as MISO systems running in parallel).
Each individual contribution is then bounded applying similar arguments  
as in \cite{Leveque-Tse-four-reg}.
 
The above mentioned five regimes derive from the fact that the dominant
contribution to ${\cal C}({\mathcal S},{\mathcal D})$ changes while varying the
system parameters. In regime I the dominant contribution is due to 
nodes lying at distance $\Theta(L)$ from the cut; in regime II the 
dominant contribution is provided by nodes which are jointly  at distance
$\omega(d_c \sqrt{\log n})$ and  $o(L)$; in regime III it is due 
to nodes at distance $O(d_c \sqrt{\log n})$ and $\Omega(d_c)$; in regime IV it is due to 
nodes at distance $o(d_c)$ and $\omega(1/\sqrt{\underline{\Phi}})$; at last,  
in regime V the dominant contribution is provided by nodes at distance  
$\Theta(1/\sqrt{\underline{\Phi}})$. A detailed derivation of the upper-bounds 
can be found in \cite{tec-rep}.

\vspace {-0mm}

\section{Lower Bounds}\label{sec:lower}
For each operational regime, it is possible to devise a communication scheme
that approaches the corresponding upper bound to within a poly-log factor.
All of our proposed schemes work as follows: first, a subset of nodes
is identified, which forms the main infrastructure through which
data is transferred across the network area.
A finite fraction of time is then assigned to the rest of the nodes to exchange traffic 
with the nodes belonging to the main infrastructure (if needed). More precisely,
time is divided into regular frames, each one comprising three phases 
of equal duration: i) an {\em access} phase, in which sources not belonging to the 
main infrastructure send data to the infrastructure; ii) a {\em transport} phase, in which data
is transferred over the infrastructure; iii) a {\em delivery} phase, in which data
is sent from the infrastructure to destinations not belonging to it. Since the {\em delivery} phase
is analogous to the {\em access} phase (by exchanging the role of transmitters and receivers), 
we will focus on the {\em access} phase only, after presenting the {\em transport} phase.

Before proceeding, we report the lower bounds obtained in \cite{Leveque-Tse-four-reg}
for homogeneous networks. Given a Homogeneous Poisson Process (HPP) 
of intensity $\psi$ over a square (or disc) of edge (radius) 
$L$, it is possible to achieve the aggregate capacity $C_n(L, \psi, \alpha)$: \vspace{1mm}
\begin{equation}\ba{rcl}
\!\!\!\!\!\! C_n(L, \psi, \alpha) &\!\!\!\!\! = \!\!\!\!\!& \left \{
\begin{array}{lc}
\!\!\omega(\bar{N}^{1 - \epsilon}) & \!\! \bar{N} \ge L^\alpha \\
\!\!\omega(\bar{N}^{2 - \epsilon} L^{-\alpha}) & \!\!  \bar{N} < L^\alpha \,,\, \alpha < 3  \\
\!\!\omega(\bar{N}^{-\epsilon} L\,\psi^{\frac{\alpha-1}{\alpha -2}} )  & \,\,   \bar{N} < L^\alpha \,,\, \alpha \ge 3 \,,\, \psi = \omega(1) \\
\!\!\omega(\bar{N}^{-\epsilon} L\,\psi^{\frac{\alpha+1}{2}})  & \!\!  \alpha \ge 3 \,,\, \psi = O(1) \\
\end{array}\right.
\label{eq:Cleveque}\ea
\end{equation}
w.h.p. for any $\epsilon > 0$. In the above expressions $\bar{N} = \psi L^2$ 
is the average number of nodes in the system.  

\subsection{\bf Transport phase}\label{sec:transport}
For what concerns the {\em transport} phase, our proposed schemes
can be considered as special cases of a general class of scheduling-routing strategies, 
according to which the network area is partitioned into cells of edge size $l$.
A cooperative multi-hop strategy is applied, in which MIMO communications are established
between the nodes belonging to neighboring cells, and global multi-hopping at the cell level is 
employed to transfer data through the network.
The proposed schemes essentially differ in: i) the subset of nodes which are used as the main 
infrastructure; ii) the chosen value of the cell edge size $l$. 
In particular, the value of $l$ allows us to classify our schemes into five main communication 
strategies (for the {\em transport} phase) which can be associated by a one-to-one correspondence to the
five operational regimes reported in Table \ref{tab:results}:
\begin{list}{\leftmargin=1em}
\item {\bf I: global MIMO}, in which $l = \Theta(L)$, and nodes employ a 
MIMO communication scheme at global network scale, without the need of cell
multi-hopping;  \vspace{-0mm}
\item {\bf II: cooperative super-cluster hopping}, in which nodes
employ a cooperative multi-hop scheme, where $l = \omega(d_c \sqrt{\log{n}})$;
\item {\bf III: cooperative inter-cluster hopping}, in which $l = \Theta(d_c \sqrt{\log{n}})$,
i.e., the cell edge size is closely related to the typical distance $d_c$ between cluster centres;
\item {\bf IV: cooperative sub-cluster hopping}, in which $l = o(d_c)$ 
and $l = \omega(1/\underline{\Phi})$, i.e., the cell edge size 
is smaller (in order sense) than the typical distance between cluster centres, 
yet the cell is large enough to allow cooperation among an increasingly number of nodes falling in it;
\item {\bf V: traditional multi-hop scheme}, in which $l = \Theta(1/\sqrt{\underline{\Phi}})$, 
and nodes resort to the traditional point-to-point multi-hop scheme,
since there is no advantage (in order sense) in employing cooperative techniques.
\end{list}
Notice that the above five strategies for the {\em transport} phase are applied to different 
infrastructures, which are selected depending on the combination of system parameters.
The basic tool that we use to extract a subset of nodes forming the main infrastructure
is a standard thinning technique, that can be applied to our class of point processes
in the sense specified by the following lemma.
\begin{lemma}\label{lemma:random}
Consider nodes ${\bf X}=  \{X\}_1^N$ 
placed according to the considered SNCP.  Then a subset of nodes
${\bf Z} \subseteq {\bf X}$ can be found w.h.p. such that ${\bf Z}$ forms a homogeneous Poisson process with intensity $\underline{\Phi}_0$, where
$\underline{\Phi}_0= g_1 \frac{n}{L^2}$ in the {\em cluster-dense}
condition and $\underline{\Phi}_0 = g_2\,q \log m \,s(d_c \sqrt{\log m})$
in the {\em cluster-sparse} condition. Here $g_1$ and $g_2$ are the
constants defined in Lemma \ref{teo:dens}.
\end{lemma}
We identify the following three main infrastructures:
\begin{list}{\leftmargin=1em}
\item {\bf dense infrastructure}, which is used in regimes I and II, but only for 
the {\em cluster-dense} condition ($\gamma < \nu/2$). In this case, we can apply Lemma
\ref{lemma:random} and extract a subset ${\bf Z}$ of cardinality $\Theta(n)$, which 
can sustain the same capacity of a homogeneous system with $n$ nodes; \vspace{-0mm}
\item {\bf clusters-core infrastructure}, which is used in regimes I,~II,~III, for 
the {\em cluster-sparse} condition ($\gamma > \nu/2$). In this case, the set
${\bf Z}$ is formed by all nodes falling within a finite distance from their
cluster centre. The cardinality of this set is still $\Theta(n)$;
\item {\bf sparse infrastructure}, which is used in regimes IV and V, for 
the {\em cluster-sparse} condition ($\gamma > \nu/2$). 
In this case, we can apply Lemma \ref{lemma:random} and extract a subset ${\bf Z}$ 
of points with density $\underline{\Phi}_0 = \Theta(n^\beta)$, where $\beta = 1-\nu -\delta(\gamma - \nu/2)$. 
The cardinality of this set is $o(n)$. 
\end{list}
Since both the dense infrastructure and the sparse infrastructure form 
a HPP, their capacity can be immediately obtained applying 
existing results for homogeneous system. The cluster-core infrastructure
is not a HPP, however it can be regarded as being uniformly dense at resolution 
higher than $d_c$. Since in regime I,II,III the cell edge size is $\Omega(d_c \sqrt{\log{n}})$,
MIMO communications between cells occur as if nodes in ${\bf Z}$ were
uniformly distributed (see \cite{tec-rep} for more details). Moreover, it can be shown that the clusters-core infrastructure can sustain the 
load due to the cooperation overhead required within each cell, but we omit the details here. 

\subsection{\bf Access phase}\label{sec:transport2}
We recall that the {\em access} phase is used by sources to inject their traffic over the
main infrastructure. Since the system capacity is ultimately determined by the
main infrastructure, the goal is to design an {\em access} phase that does
not constitute a system bottleneck, while at the same time inducing a 
uniform traffic matrix over the main infrastructure.
These design principles led us to select the following three {\em access} strategies:
\begin{list}{\leftmargin=1em}
\item {\bf SISO access scheme}. This is the simplest strategy, and it is used to access 
the dense infrastructure. In this case, it is sufficient to employ a single-hop 
point-to-point transmission between each source and one of the closest nodes belonging 
to ${\bf Z}$, thanks to the fact that the network is almost uniformly dense; \vspace{0mm}
\item {\bf SIMO access scheme}. This is used to access the nodes of the
clusters-core infrastructure, employing a SIMO technique similar to the 
relaying scheme proposed in \cite{shah}\footnote{In \cite{shah}, authors present a technique that allows
nodes located in low-density areas to relay their data over
densely populated areas, by exploiting the diversity gain intrinsically available
in high-density regions thanks to the presence of many nodes acting
as an array of receiving antennas.};
\item {\bf hierarchical access scheme}. This is used to access the nodes of the
sparse infrastructure, and required us to develop a novel scheduling-routing 
strategy specifically tailored to this case.
\end{list}
Due to lack of space, we restrict ourselves to a brief description of the hierarchical access scheme,
which is the most intriguing one\footnote{The interested reader is referred 
to \cite{tec-rep} for a detailed description and analysis of all {\em access} schemes.}. 
In this case, traffic produced within highly dense
regions of the network area (e.g., the clusters cores in Figure \ref{subfig:example3})
needs to be gradually spread out through a sequence of intermediate, local transport infrastructures
nested one within the other,
This construction is needed both to avoid the formation of local bottlenecks around the
cluster centres, and to evenly balance the traffic towards the node of the main infrastructure. 
Intermediate transport infrastructures are obtained by applying the thinning technique of 
Lemma \ref{lemma:random} within certain domains (specified later), surrounding the clusters' centres, 
nested one within the other.  
To simply and effectively  balance the traffic 
data are delivered within each local infrastructure to  randomly destination
 nodes.

The sequence $k = 0,1,\ldots, K_{\max}$ of nested domains is carefully chosen in such a way that:
i) the first domain in the sequence coincides with the network area, hence the
corresponding infrastructure is the main transport infrastructure of the network, of density $\underline{\Phi}$,
which is shared by all data flows;
ii) the infrastructure extracted in each domain $k > 0$ can pass to the infrastructure of 
domain $k-1$ all traffic generated by nodes contained in it;
iii) the total number of domains grows at most like $\log{n}$.

\tgifps{6}{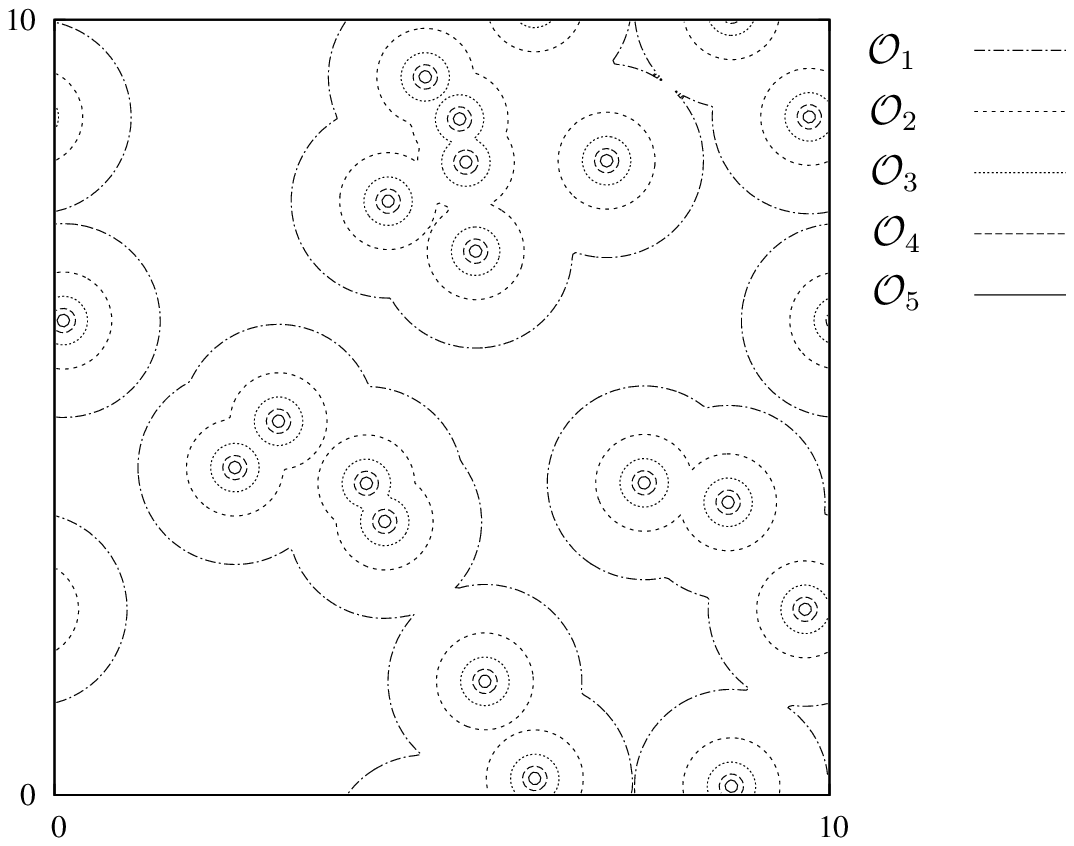}{Example of construction of nested domains ${\cal O}_k$ for the 
topology depicted in Figure \ref{subfig:example3}. Domain ${\cal O}_1$ is characterized by 
$d_1 = 0.5 d_c$.}

Conditions i) and ii) guarantee that the system capacity is throttled by the 
lowest infrastructure (the main transport infrastructure) and no bottleneck arises
within any higher infrastructure. 
Condition iii) guarantees that, even if we devote to each layer-$k$ infrastructure 
the same  fraction of time, the total overhead due to the {\em access} phase causes at most a $\log n$ loss in the overall system capacity.  
 
We now specify one possible way to jointly achieve the three conditions above.
We build a sequence of nested domains ${\cal O}_k$, $k = 0,1,2\ldots, K_{\max}$, as follows.
The first domain is ${\cal O}_{0}={\cal O}$, meeting condition i). 

For the generic point $\xi \in {\cal O}$, let $d_{\min}(\xi)=\min_j ||\xi- c_j||$ be
the distance between $\xi$ and the closest cluster centre.
We define domains ${\cal O}_k$, for $k\ge 1$, as follows:
${\cal O}_k=\{\xi\in {\cal O}:  d_{\min}(\xi) \le d_k\}$, where $d_k$ are
a set of decreasing distances, i.e., $d_1 > d_2 > \ldots > d_{K_{\max}}$. 
Domain ${\cal O}_k$ is, in general, composed of a random
number $J_k$ of disjoint regions ($J_k \le M$), corresponding to the connected
components of the standard Gilbert's model of continuum percolation \cite{meester}
with ball radius $d_k$. Figure \ref{fig:distanze} shows examples of domains ${\cal O}_k$
having different values of $d_k$. Let $\{{\cal  I}_k^j\}_{j}$ be the set of disjoint
regions ($1 \le j \le J_k$) forming domain ${\cal O}_k$.

We set the largest $d_k$, namely $d_1$, equal to $d_1 = \mu\,d_c$, where $\mu$ is a small constant.
Choosing $\mu$ sufficiently small, in such a way that the associated
Gilbert's model is below the percolation threshold (we need $\mu < \mu^*$, where
$\mu^* \approx 0.6$), we have the property that
the maximum number of clusters centres belonging to the same
region ${\cal  I}^j_1$ is $O(\log n)$ w.h.p. \cite{meester}.
Since by construction ${\cal O}_{k+1} \subset {\cal O}_{k}$, the same
property holds for all $k>1$. It follows that, in terms of physical extension, 
the area $|{\cal  I}^j_k|$ of region ${\cal  I}^j_k$ lies w.h.p. in the interval 
$\pi d_k^2 \le |{\cal  I}^j_k| \le  \pi d_k^2 \log n$.

We further observe that the density of nodes at any point within ${\cal O}_k$ ($k\ge 1$) 
can be lower bounded by $\lambda_k =  q\,d_k^{-\delta}$,
by considering the contribution of the closest cluster centre only. 
Hence, it is possible to extract from ${\cal O}_k$ ($k\ge 1$) 
a set of points ${\bf Z}_k$ forming a HPP with intensity $\lambda_k$. 
Note that in the domain ${\cal O}_0$ we have $\lambda_0 =  \underline{\Phi}$. 
Distances $d_k$, for $k \ge 2$, are then assigned in such a way that 
$\lambda_k = 2^{k-1} \lambda_1$, i.e., the intensities of the nested
transport infrastructures form a geometric progression.
This requires to set $d_k = d_1 2^{-\frac{k-1}{\delta}}$.
Since the maximum node density in the network is \mbox{$\overline{\Phi} <  G_2 q\,\log{m}$} (see Lemma \ref{teo:dens}),
we have $K_{\max} = 1 + \lfloor \log_2(q \log{m}/\lambda_1)\rfloor = O(\log{n})$, hence the total number of domains 
satisfies condition iii). 

It remains to show that each domain $k < K_{\max}$ can receive the traffic generated by
domain $k+1$. To this purpose, we need to show that each region ${\cal  I}^j_k$
can handle the traffic produced by all components of domain $k+1$ nested in it.
Let ${\cal H}^j_{k+1}$ be the set of indexes $h$ of regions ${\cal  I}^h_{k+1}$ 
falling in ${\cal  I}^j_k$. 
Moreover, let $M^j_k$ be the number of cluster centres falling within ${\cal  I}^j_k$.

The area of ${\cal  I}^j_k$ can be expressed as $|{\cal  I}^j_k| = M^j_k \pi d_k^2 \zeta_k$,
where $\zeta_k < 1$ is a reduction factor that accounts for the overlapping among the 
discs of radius $d_k$ forming region ${\cal  I}^j_k$. The sum of the areas of all nested
regions ${\cal  I}^h_{k+1}$ is instead given by $\sum_{h \in {\cal H}^j_k}|{\cal  I}^h_{k+1}| = M^j_k \pi d_{k+1}^2 \zeta_{k+1}$,
where $\zeta_{k+1} > \zeta_{k}$ because the degree of overlapping among the 
discs reduces for decreasing values of $d_k$. 
Since $(d_k/d_{k+1})^2 = 2^{2/\delta}$, we conclude that the ratio between $|{\cal  I}^j_k|$
and $\sum_{h \in {\cal H}^j_k}|{\cal  I}^h_{k+1}|$ is bounded. 
This is important, as it allows to exploit to full capacity of the infrastructure 
extracted in ${\cal  I}^j_k$ to spread out the traffic coming from 
nested regions ${\cal  I}^h_{k+1}$ over the larger region ${\cal  I}^j_k$.

Moreover, using the expressions \equaref{Cleveque} it can be shown that the
aggregate capacity of nested regions ${\cal  I}^h_{k+1}$ is larger than the capacity of
region ${\cal  I}^j_k$. This allows to conclude that domain ${\cal O}_0$ (i.e., the main
infrastructure) acts as the system bottleneck.
Indeed, the number of points in ${\cal  I}^j_k$ is $$ M^j_k \pi \lambda_k d_k^2 \zeta_k = M^j_k \pi \lambda_1 d_1 2^{(k-1)(1-\frac{2}{\delta})} \zeta_k$$
The total number of points in regions  ${\cal  I}^h_{k+1}$ has the same expression, substituting $k$ with $k+1$.
Since $\delta > 2$, and $\zeta_{k+1} > \zeta_{k}$, the total 
number of points in regions ${\cal  I}^h_{k+1}$ is larger.
This guarantees that the aggregate capacities of the nested infrastructures
is higher than the capacity of ${\cal  I}^j_k$ in the first regime of \equaref{Cleveque}, in which $C_n = \omega(\bar{N}^{1-\epsilon})$.

In the third regime of \equaref{Cleveque}, the capacity (either of region ${\cal  I}^j_{k}$ or
the aggregate capacity of nested regions ${\cal  I}^h_{k+1}$) would be proportional to   
$2^{k[\frac{\alpha-1}{\alpha-2} - \frac{1}{\delta} - \epsilon(1 - \frac{2}{\delta})]} \zeta_k$.
Since $\frac{\alpha-1}{\alpha-2} > 1 > \frac{1}{\delta}$, and $\epsilon$ is small, the
capacity increases with $k$. At last, in the forth regime of \equaref{Cleveque} the capacity would be
proportional to $2^{k[\frac{\alpha+1}{2} - \frac{1}{\delta} - \epsilon(1 - \frac{2}{\delta})]} \zeta_k$,
which again increases with $k$. One can verify that capacities still form a non-decreasing sequence when we change 
regime passing from layer $k$ to layer $k+1$. 

We conclude that the chosen sequence of nested local infrastructures satisfies
the conditions that allow to balance the traffic towards the nodes
of the main infrastructure at most with 
a $\log n$ penalty factor to the system capacity. 

\section{conclusions}
We have characterized the asymptotic capacity of networks
whose nodes are distributed according to a doubly stochastic shot-noise Cox process.
This point process provides an interesting, analytically tractable model of clustered 
random networks containing large inhomogeneities in the node density. 
Our study has revealed the existence of additional operational regimes
with respect to those identified in previous work, and the need of novel  
scheduling and routing strategies, specifically tailored to each regime,
to approach the maximum system capacity.

\end{sloppypar}
\end{document}